# FogGIS: Fog Computing for Geospatial Big Data Analytics

Rabindra K. Barik[1], Harishchandra Dubey[2], Arun B. Samaddar[3], Rajan D. Gupta[4], Prakash K. Ray[5]

[1]School of Computer Application, KIIT University, Bhubaneswar, India
rabindra.mnnit@gmail.com

[2]Electrical Engineering, The University of Texas at Dallas, USA
harishchandra.dubey@utdallas.edu

[3]Director, NIT Sikkim, India
absamaddar@yahoo.com

[4]Civil Engineering Department, MNNIT Allahabad, India
gupta.rd@gmail.com

[5]Electrical Engineering Department, IIIT Bhubaneswar, India
prakash@iiit-bh.ac.in

*Abstract*— Cloud Geographic Information Systems (GIS) has emerged as a tool for analysis, processing and transmission of geospatial data. The Fog computing is a paradigm where Fog devices help to increase throughput and reduce latency at the edge of the client. This paper developed a Fog Computing based framework named FogGIS for mining analytics from geospatial data. It has been built a prototype using Intel Edison, an embedded microprocessor. FogGIS has validated by doing preliminary analysis including compression and overlay analysis. Results showed that Fog Computing hold a great promise for analysis of geospatial data. Several open source compression techniques have been used for reducing the transmission to the cloud.

## I. Introduction

Geographic Information System (GIS) is a system of software and computer hardware that enables end-users to retrieve, store, and analyze huge amount of geospatial data from a various sources [1]. GIS is applied in decision making, storage of various kinds of data, bringing data and maps to a common scale as per the user needs, superimposing, querying and analyzing the data and designing/ presenting final maps/ reports to the administrators and planners [2]. The utility of GIS for planning of land resources and decision making has become widely popular and are being used for a wide range of applications. GIS has emerged as a powerful tool in integrating and analyzing various thematic layers along with their attribute information to create and visualize alternative planning scenarios for planners and decision makers. The user friendliness of GIS is a feature that has made GIS a preferred platform for planning all over the world, coupled with various analysis and modelling functionalities.

GIS can play an important role in various applications such as environmental monitoring, natural resource management, healthcare, land use planning and urban planning. GIS integrates common database operations such as query formation, statistical computations and overlay analysis with unique visualization and geographical functionalities.

These characteristics distinguish GIS from other information systems and make it valuable to a wide range of public and private enterprises for explaining events, predicting outcomes and designing strategies. The GIS technology and cloud computing has been merged to perform a value added services that give rise to geospatial cloud computing. The geospatial data have rich information about temporal as well as spatial distributions. In traditional setup, we send the data to the cloud where these are going for further processing and analysis.

The Fog Computing provides low-power gateway that can increase throughput and reduces latency near the edge of the geo-spatial clients. It reduces the storage needed for geospatial big data in the cloud. In addition, reduction in the required transmission power results in overall improvement in efficiency. Fog devices can act as a gateway between clients such as mobile phones [22]. In this paper, we let the geospatial data be processed at the edge using Fog computing device. The present paper made the following contributions to the GIS systems:

- FogGIS framework is proposed for improved throughput and reduced latency for analysis and transmission of geospatial data
- Intel Edison was employed as the fog computing device
- Various compression techniques were used for reducing the data size, thereby reducing transmission power
- Geospatial data analysis scheme and overlay analysis in thin clients environment was performed using FogGIS framework. It has been performed a case study by doing overlay analysis of city of Alaska, USA

## II. RELATED WORKS

### A. Geospatial Cloud

Cloud computing provides adequate storage and computational infrastructure for implementation of geo-spatial analysis prototypes. This model provides a transition from PC to cloud servers. Cloud computing and other web processing architectures creates an open environment in web with shared assets [5-7].

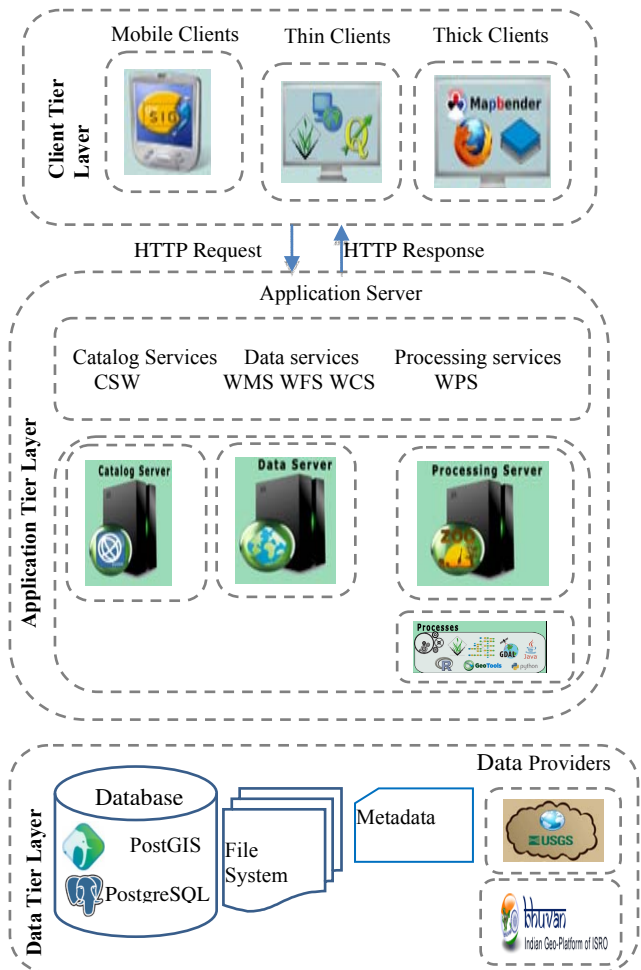

Fig. 1. System architecture for Geospatial Cloud Model [10].

Geospatial Cloud delivers a platform in which organizations interrelate with technologies, tools and expertise to nurture deeds for producing, handling and using geographical statistics and data. Likewise, Geospatial Cloud deploy a unique-instance, multitenant design and permitting more than one client to contribute assets without disrupting each other. This integrated hosted service method helps installing patches and application advancements for user's transparency. It's another characteristic is embrace of web services and as an established architectural methodology in engineering [8-9]. Many cloud platforms uncover the applications statistics and functionalities via web service. These permit clients to query/update different types of cloud services and applications data programmatically, along with the provision of a typical tool to assimilate different cloud applications in the software cloud with enterprise SOA infrastructure. Figure 1 shows the system architecture for Geospatial Cloud Model adapted from [10].

The client tier layer consists of thick clients, thin clients and mobile clients with visualization functionality for spatial information. Mobile clients are users operating through mobile devices. The users those are working on web browsers are defined to be thin clients. In thin clients, users do not require any additional software for the operation. Thick clients are the users processing or visualising the spatial data in standalone system where it requires additional software for full phase operation.

The Application Tier comprises the main geo-spatial services executed by servers. It enables intermediate amongst the different clients and providers. In top of the application tier, dedicated server for application has been operated for different services i.e. Web Map Service (WMS), Web Coverage Service (WCS), Web Feature Service (WFS), Web Catalog Service (CSW) and Web Processing Service (WPS). The dedicated application server is responsible for requests to and response from client to application server. In addition, application services include three types of server application i.e. catalog servers, data servers and processing servers. Catalog severs are used to search the metadata information regarding the stored spatial data. Catalog server is one of the important system components for controlling spatial information in cloud environment. In the catalog service, a standard publish-find-bind service framework are implemented which has been defined by OGC web service architecture. Data server deals with the WMS, WCS and WFS [11].

Processing server offers a geospatial processes which allows different clients to smear in WPS standard spatial data. The detail explanation of every processes done by client request, forward the desire processing service with input of several factors, specifies and provides definite region in leaping box and feedbacks with composite standards. Data tier Layer comprises of the various data in spatial form and related info. System utilizes the layer to store, recover, manipulate and update the spatial data for further analysis. Data providers can be store in different open source DBMS packages, simple file system or international organizations (e.g., Bhuvan, USGS). It has been shown from the system architecture of Geospatial Cloud that geospatial data are one of the key components in data layer for the handling of huge amount of data in terms of various spatial analysis. The amount of data which has been handling in Geospatial Cloud computing, it requires geospatial data from the various components. That gives rise to the concept of geospatial big data aspects and that will discuss in the next section.

### B. Geospatial Big data

Big data are data those distribution, scale, diversity or timeliness needs the employ of new robust technological architectures and data analytics to enable or permit insights that unlock new source of business value. Big data typically includes variety of data sets with variable sizes ahead of the ability of generally used software tools to manage, capture, curate and process the data set within a acceptable elapsed time [12]. Big data can come in multiple forms. Most of the big data is semi-structured, Quasi structured or unstructured, which requires numerous techniques and tools to analyze and process. Analysis of big data sets can discover the new correlations to spot business trends, combat crime, and prevent diseases.

vector data. Graph data appear in the form of road networks. Here, an edge represents a road segment and a node represents an intersection or a landmark.

There are various regions behind the disadvantageous of geospatial cloud computing with geospatial big data. As we know reliability, manageability and cost saving, are the key factors in which cloud computing always be one of advantageous over other emerge technology for data processing. But in terms of security and privacy are the main concerns for the processing of sensitive data. Particularly in

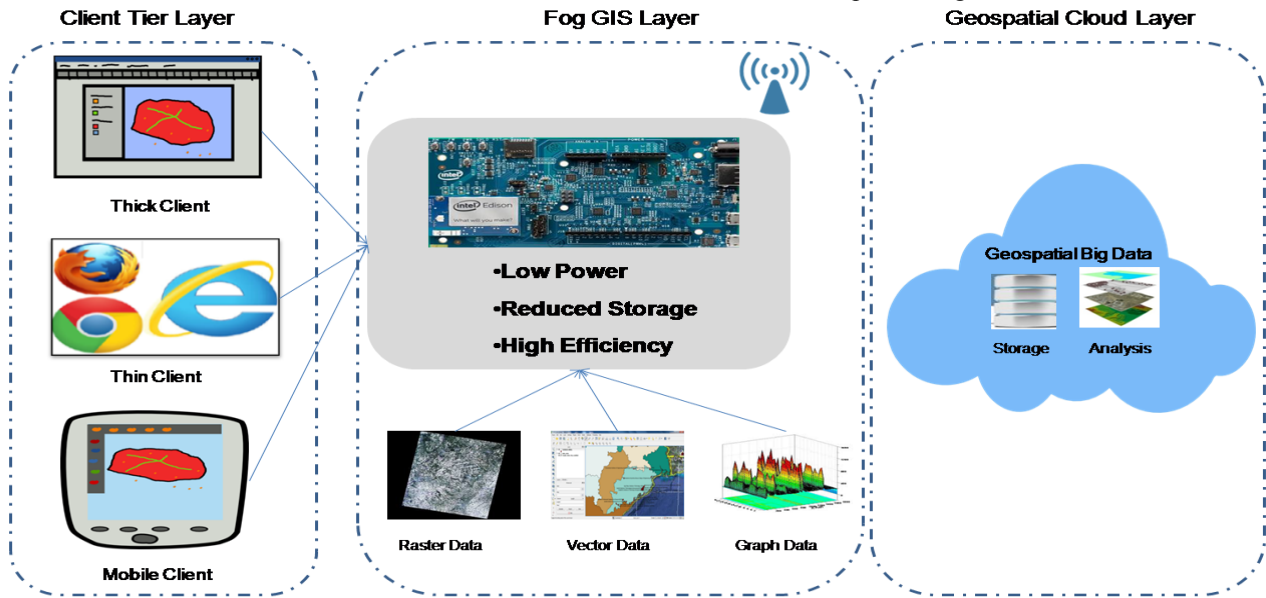

Fig. 2. Conceptual diagram of the proposed FogGIS framework for power-efficient, low latency and high throughput analysis of the geospatial big data.

Big data sets are growing rapidly because they are increasingly gathered by economical and numerous radio-frequency identification (RFID) readers, information sensing mobile devices, cameras, microphones, wireless sensor networks, aerial (remote sensing) and software logs. Geospatial data has always been big data with the combination of remote sensing, GIS and GPS data [13]. In these days, big data analytics for geospatial data are getting considerable attention to allow users to analyze huge amounts of geospatial data. Geospatial big data usually refers to spatial data sets beyond the capacity of present computational environment.

Generally, geospatial data has been categorized into raster data, vector data and graph data. Raster data include geospatial images which are obtained by satellites, security cameras and aerial vehicles. The raster data has been provided by different government agencies for using in various analyses. It can be extract number of feature from these raster data. Change detection and pattern mining are two examples in which data analyst does. Vector data consist of points, lines and polygons features. For examples, in Google map, the various temples, bus stops and churches has been marked thorough points data whereas lines and polygons corresponds to the road networks. Spatial correction pattern analysis and hot spot detection are the analysis which can be done through health geoinformatics scenario, data are so sensitivity for further processing and analysis [14]. Thus, for minimization of privacy and security risks, it has to be used as per the user context for limited amount of data access within the limited framework. After processing within the limited framework, it will transfer to the next level for the final processing of data analysis. That wills benefits for data security and privacy. Thus, fog computing comes into picture for geospatial big data processing in our present study.

*C. Fog Computing*

Fog computing was coined by Cisco in 2012 [15]. It refers to a computing paradigm that uses interface kept close to the devices that acquire data. It introduces the facility of local processing leading to reduction in data size, lower latency, high throughput and high power efficiency of the cloud-based systems. It has been implemented on smart cities development [16] and healthcare [17]. The Fog computing have been successfully used in healthcare to translate the speech therapy from clinic to home [18-20]. The Fog devices are embedded computers such as Intel Edison that acts a gateway between cloud and mobile devices such as smart phones and mobile GIS.

## III. FOGGIS FRAMEWORK

This section describes various components of the proposed FogGIS framework and discusses the methods implemented in it. We discuss the hardware, software and methods used for compression of geospatial big data.

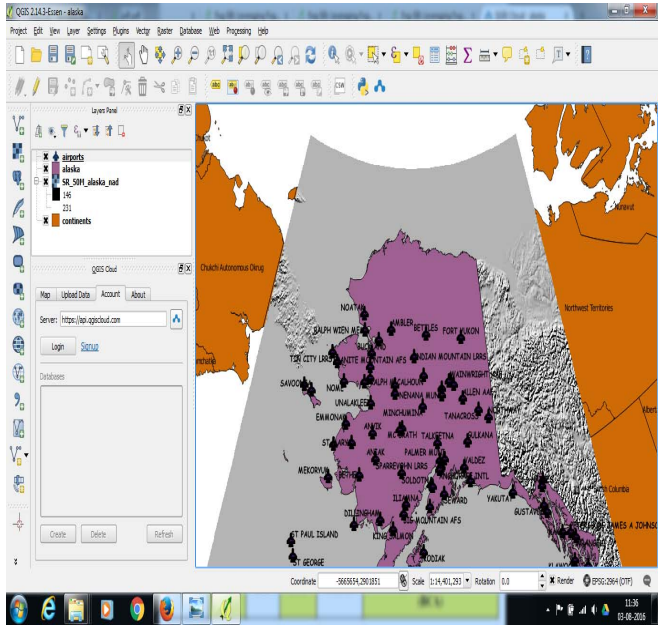

Fig. 3. Overlay operation on thick client environment in FogGIS framework.

### A. Intel Edison

We employed Intel Edision as Fog computing device in proposed FogGIS framework [21]. Intel Edison is powered by a rechargeable lithium battery. It contains dual-core, dual-threaded 500MHz Intel Atom CPU along with a 100MHz Intel Quark microcontroller. It possesses 1GB memory with 4GB flash storage. It supports IEEE 802.11 a,b,g,n standards and can connect to WIFI. It has been used UbiLinux operating system for running compression utilities.

Figure 2 shows the proposed FogGIS framework. The fog device acts as a gateway between thick, thin and mobile clients and cloud layer. The proposed FogGIS framework has three layers as client tier layer, geospatial cloud layer and FogGIS layer. In client tier, the categories of users have been further divided into thick client, thin client and mobile client environment. Processing of geospatial data can be possible within these three environments. Geospatial Cloud layer is mainly focused on overall storage and analysis of geospatial data. The Fog layer works as middle tier between client tier layer and geospatial cloud layer. It has been experimentally validated that the Fog layer is characterized by low power consumption, reduced storage requirement and overlay analysis capabilities.

### B. Lossless Compression Techniques

In the present study, we have a number of popular compression algorithms for reducing the data size in fog layer. The concept of compression in GIS is not new, it have been used in network GIS and mobile GIS [23-25]. In this paper, we translated the compression from mobile GIS to fog layer [20]. The geo-spatial data is compressed on the Fog computer that later transmits the data to cloud layer. The cloud layer has the choice to store the compressed data or decompress it before processing, analysis and visualization. We used only lossless techniques in this paper such as .zip, .tar.gz, .gzip.

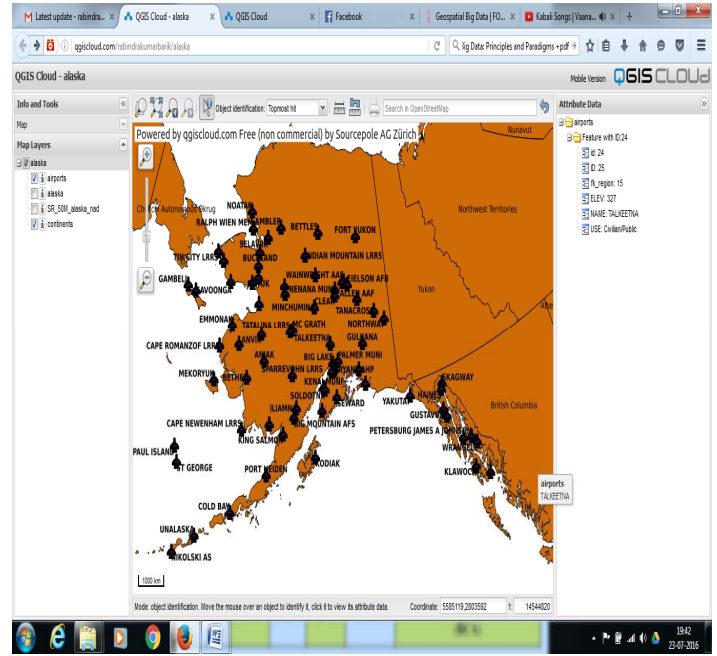

Fig. 4. Overlay operation on thin client environment in FogGIS framework.

The results have been obtained by using various lossless compression techniques done at the Fog gateway which has shown in Table I

### C. Geospatial Analysis of Alaska city, USA

In this section, data analysis particularly overlay analysis is performed for city of Alaska, USA. Overlay Analysis is one of the important data analysis in which we superimpose various geospatial data in a common platform for better analysis of raster and vector geospatial data. We performed the case study on the city of Alaska, USA. We downloaded the freely available dataset both raster and vector geospatial data[26]. It has been found that one SRTM raster data and three number of vector data of Alaska in EPSG:2964 file format. Continents boundary, City boundary of Alaska and airport location details of Alaska are the three number of vector data have been used; which are in shape file formats. The overlay analysis of various vector data and raster data of particular area has been performed. Initially, the downloaded datasets have been opened with Quantum GIS; desktop based GIS analysis tools, and performed some join operations which has been shown in Figure 3.

The desired overlay operation has been done with standalone application, are known as thick client operation. In Quantum GIS, plugin named as QGISCloud has been installed. The said plugin has the capability of storing various raster and vector data set in cloud database for further overlay analysis.

After storing in cloud database, it also generates the mobile and thin client link for visualization of both vector and raster data set. Figure 4 shows the overlay operation on thin client environment. The Figure 3 and 4 shows the overlay analysis on thick and thin client respectively. We can see that the overlay analysis is a useful technique for visualization of geospatial data.

TABLE I. PERFORMING COMPRESSION ON FOG GIS FRAMEWORK USING GLOBAL MAP DATA[28].

| Geo-spatial Data | Original Data Size (in MBs) | .tar.gz Compressed Size (in MB) | .iso Compressed Size (in MB) | .zip Compressed Size (in MB) | .tar Compressed Size (in MB) | .gzip Compressed Size (in MB) | .zipx Compressed Size (in MB) |
|---|---|---|---|---|---|---|---|
| Coast Line-Shapefile | 6.7 | 4.9 | 5.2 | 5.1 | 6.2 | 5.8 | 5.6 |
| Coast Line-Geodatabase | 3.2 | 2.7 | 2.9 | 3.2 | 3.4 | 3.4 | 3.6 |
| Political Boundaries Areas-Shapefile | 47.3 | 33.7 | 33.6 | 33.4 | 32.8 | | |
| Political Boundaries Areas-Geodatabase | 19.7 | 17.3 | 16.4 | 16.2 | 16.0 | 15.8 | 15.6 |
| Political Boundaries Lines-Shapefile | 47.5 | 19.6 | 24.5 | 25.2 | 24.6 | 24.4 | 26.2 |
| Political Boundaries Lines-Geodatabase | 21 | 10.5 | 12.7 | 11.8 | 13.9 | 14.6 | 14.4 |
| Canals and Aqueducts-Shapefile | 2.1 | 1.1 | 1.2 | 1.5 | 1.7 | 1.8 | 1.9 |
| Canals and Aqueducts-Geodatabase | 1.5 | 0.932 | 0.942 | 0.938 | 0.936 | 0.939 | 0.942 |
| Inland Water Areas-Shapefile | 49.1 | 33.2 | 36.2 | 34.6 | 35.3 | 34.4 | 36.4 |
| Inland Water Areas-Geodatabase | 20.5 | 18.2 | 18.4 | 18.6 | 18.8 | 19.2 | 19.0 |
| Water Courses—Shapefile | 345.7 | 330.7 | 332.7 | 333.7 | 331.7 | 333.8 | 333.2 |
| Water Courses—Geodatabase | 163.9 | 105.1 | 111.9 | 110.4 | 110.6 | 110.2 | 111.8 |

We used the global map data for benchmarking the various compression algorithms [25]. The Table I shows the compressed data size and original data size for various compression procedures. The compression procedures used are .tar.gz, .iso, .zip, .tar, .gzip, .zipx. Clearly, the compression ratio depends on the data type and size. However, the .tar.gz has consistently performed the best in terms of compression ratio for Global Map Data [29].

IV. CONCLUSIONS

In this paper, we developed and validated FogGIS framework that employed Fog gateway in a cloud GIS model. Intel Edision processor was used as the fog computer. The Fog gateway reduces the storage space requirments, transmission power, increased throughput and reduced latency leading to overall efficiency of GIS system using FogGIS as an intermediate gateway. The FogGIS framework introduces edge intelligence in geospatial cloud environment. In future, we would like to add more intelligent processing at the Fog layer in mobile client environments.